\documentclass[sigconf]{acmart}

\usepackage{array}
\usepackage{enumitem}
\newcolumntype{L}[1]{>{\raggedright\arraybackslash}p{#1}}

\AtBeginDocument{}

\setcopyright{none}
\renewcommand\footnotetextcopyrightpermission[1]{}
\acmConference[NSAD '26]{Anonymous Submission to NSAD 2026}{2026}{Under Review}
\acmYear{2026}
\acmDOI{}

\begin{document}

\title{BMOA: Baseline--Mechanism--Outcome Attribution for Compiler-Induced Numerical Deviations}
\author{%
Hailong Jiang\textsuperscript{1,*},
Emran Hossain\textsuperscript{1},
Feng Yu\textsuperscript{1},
Chunwei Xia\textsuperscript{2},\\
Mengfei Ren\textsuperscript{3},
Jianfeng Zhu\textsuperscript{4},
Qiang Guan\textsuperscript{4}\\
\textsuperscript{1}Youngstown State University, USA \quad
\textsuperscript{2}University of Leeds, UK\\
\textsuperscript{3}Baylor University, USA \quad
\textsuperscript{4}Kent State University, USA\\
\texttt{\{hjiang,fyu\}@ysu.edu},
\texttt{ehossain@student.ysu.edu},
\texttt{C.Xia@leeds.ac.uk}\\
\texttt{Mengfei\_Ren@baylor.edu},
\texttt{\{jzhu10,qguan\}@kent.edu}\\
\scriptsize
\textsuperscript{*}Corresponding author
}
\renewcommand{\shortauthors}{Hailong Jiang etc.}

\begin{abstract}
Formalizing compiler-aware numerical correctness requires distinguishing what an observed floating-point difference means, what compiler behavior the evidence supports, and what numerical consequence follows. Existing testing workflows often collapse these questions into a pass/fail mismatch. We introduce Baseline--Mechanism--Outcome Attribution (BMOA), a diagnostic framework that separates the comparison relation and system boundary, the evidence-supported compiler mechanism, and the reference-qualified accuracy outcome. BMOA combines operational strict floating-point, transformation-local, reproducibility, cross-compiler, and higher-precision comparisons, while preserving mixed, ambiguous, and unknown attributions when evidence is insufficient. Each record retains inputs, configurations, numerical metrics, and supporting artifacts for audit. We evaluate BMOA on six scientific-computing kernels, deterministic stress-input families, and controlled Clang configurations on ARM64. A 1,276-record attribution corpus and a 162-instance controlled mechanism matrix show that baseline choice changes diagnoses, compiler-induced deviation does not imply accuracy loss, and cancellation and large dynamic range expose the strongest effects within the targeted matrix. BMOA converts raw mismatches into explicit, auditable, evidence-bounded records. Although it is not itself a proof system, these records provide an empirical foundation for future formal specifications and proof obligations for compiler-aware numerical correctness.
\end{abstract}


\maketitle

\section{Introduction}
\label{sec:intro}

Formal reasoning about compiler behavior is only as precise as the correctness
claim being formalized. For floating-point programs, preserving an operational
evaluation path, matching a numerical output, and improving accuracy relative
to a numerical specification are distinct claims. Compiler testing commonly
compares outputs across compilers, optimization levels, or program
variants. IEEE~754 arithmetic is rounded and order-sensitive
\cite{ieee754,goldberg1991}, while compilers may reassociate expressions,
contract multiply-adds, vectorize reductions, or apply relaxed fast-math
assumptions \cite{monniaux2008,notzli2016lifejacket,becker2019icing}. Comparing
an optimized execution with an operational strict-FP configuration asks whether
the selected evaluation behaviors differ. Comparing either result with a
higher-precision reference asks how accurately it approximates that reference.
Neither observation implies the other. Without separating them, a test---or a
later proof obligation---can mistake an evaluation-path deviation for accuracy
loss or treat reference proximity as evidence of compiler causality.

Several systems are proposed to detect or localize floating-point variability. FLiT detects and
bisects inconsistent results across compiler settings and
platforms~\cite{sawaya2017flit}, pLiner isolates source
lines~\cite{guo2020pliner}, Ciel isolates expressions in heterogeneous
code~\cite{miao2023ciel}, and Varity quantifies cross-platform
variation~\cite{laguna2020varity}. Input generators further search for cases
that expose compiler-induced variability
\cite{yu2023inputs,miao2024ranges}. These advances answer \emph{whether} and
\emph{where} outputs differ. What remains is a claim-and-evidence interface:
which comparison gives the difference meaning, which compiler behavior is
supported by the available evidence, and which numerical specification
determines its consequence. Such an interface is needed both for empirical
diagnosis and for formulating well-scoped correctness properties and proof
obligations.

We introduce \emph{Baseline--Mechanism--Outcome Attribution} (BMOA) to provide
this interface. For an observed difference, BMOA emits a structured
$\langle B,M,O\rangle$ record. The baseline $B$ identifies the comparison
relation and system boundary that give the observation meaning; the mechanism
$M$ identifies a compiler or execution category supported by controlled
evidence; and the outcome $O$ reports improvement, degradation, or no change
relative to a named numerical reference. Mixed, ambiguous, and unknown labels
make unresolved evidence explicit rather than inventing a cause. BMOA is not
itself a proof system. Instead, it provides an empirical foundation from which
future formal specifications can identify the relevant comparison relation,
compiler assumptions, numerical objective, and associated proof obligations.

We evaluate BMOA in one controlled CPU workflow with two coordinated
components. The \emph{attribution corpus} contains 1,276 baseline-specific
records from five common kernels and deterministic stress inputs. The
\emph{controlled mechanism matrix} adds a purpose-built FMA kernel and contains
162 configuration instances spanning nine kernel--input pairs, nine
configurations, and two Clang distributions. Its ablations, repeated runs,
Decimal validation, and cross-distribution comparisons support the
attributions. The observed improvements and degradations show that strict-FP
deviation does not necessarily imply accuracy loss.

This paper makes three contributions:\\

\begin{itemize}[label=\tiny$\blacksquare$, leftmargin=0pt]
  \item It formulates post-detection diagnosis as BMO attribution, separating
  comparison meaning, evidence-supported mechanism, and reference-relative
  numerical outcome.
  \item It defines an evidence-combination procedure that preserves mixed,
  ambiguous, and unknown attributions instead of inferring causality from a
  flag or instruction alone.
  \item It evaluates BMOA across kernels, sensitive inputs, compiler
  configurations, numerical references, and repeated runs, exposing when
  compiler-induced changes improve or degrade accuracy.
\end{itemize}

\section{Diagnostic Model}
\label{sec:model}

\subsection{Baselines Define Meaning}

Let $y_s(x)$ be the output of an operational strict-FP configuration, $y_c(x)$
the output of candidate configuration $c$, and $r(x)$ a higher-precision
reference. For a distance $d$, BMOA distinguishes
\begin{equation}
  \begin{aligned}
  D_s(x)&=d(y_c(x),y_s(x)),\\
  E_i(x)&=d(y_i(x),r(x)),\qquad i\in\{s,c\}.
  \end{aligned}
  \label{eq:distances}
\end{equation}
$D_s>0$ establishes an observable difference under the two configurations; it
does not identify the more accurate result. Comparing $E_c$ with $E_s$ answers
that separate question. Thus $D_s>0$ can coexist with $E_c<E_s$, while
$D_s=0$ does not imply $E_s=0$. This distinction avoids using a strict result
as an accuracy oracle, a pitfall in floating-point validation
\cite{monniaux2008,becker2018combining}.

BMOA also uses a \emph{mechanism-local} baseline to compare configurations
that differ in a targeted compiler behavior, a \emph{reproducibility} baseline
to compare repeated executions of one binary, and an \emph{across-compiler}
baseline to compare corresponding configurations. These comparisons are not
interchangeable: reproducible and cross-platform floating-point computation
requires its own evidence and algorithms
\cite{demmel2013repro,collange2015repro}.

\subsection{Mechanisms Are Evidence-Bounded}

A compiler option may activate several transformations, and an instruction may
be present without affecting the reported output. BMOA therefore attributes a
\emph{mechanism category}, not a unique compiler pass or instruction. A label
is supported when a controlled pair changes the output and corroborating
code-generation evidence is consistent with that category. Multiple supported
categories produce a \emph{mixed} label; an inseparable compound intervention
produces \emph{ambiguous}; and missing support produces \emph{unknown}. Outcome
is assigned independently, since one mechanism may improve one input and
degrade another.

\section{Method}
\label{sec:method}

\subsection{Workflow}

\begin{figure*}[t]
  \centering
  \includegraphics[width=0.9\textwidth]{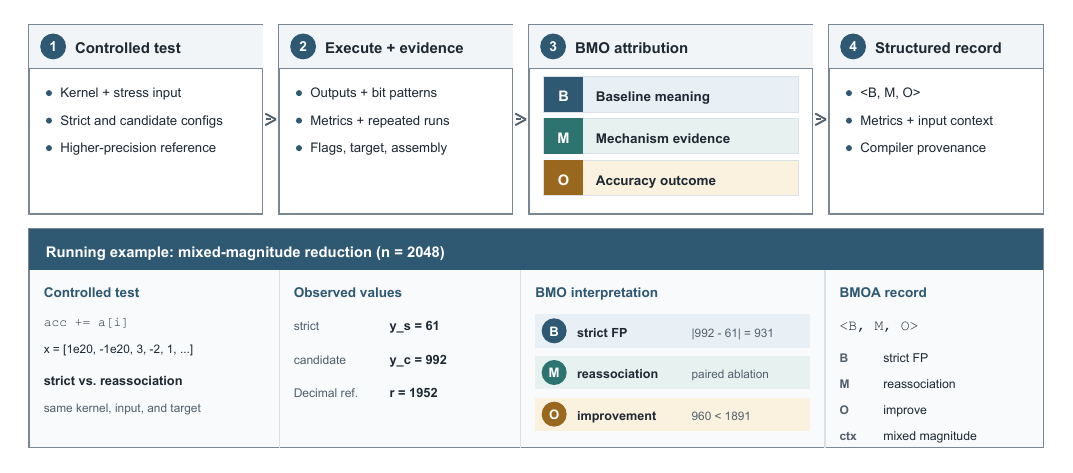}
  \caption{BMOA workflow and running example. The mixed-magnitude reduction
  differs from strict FP, the paired ablation supports reassociation, and the
  candidate is closer to the Decimal reference. The three claims use different
  comparisons and therefore remain separate fields.}
  \Description{Four-stage BMOA workflow followed by a running reduction example with baseline, mechanism, and outcome evidence.}
  \label{fig:bmoa-overview}
\end{figure*}

Figure~\ref{fig:bmoa-overview} summarizes the workflow. We define a controlled
test as $T=(P,x,C,r)$, comprising kernel $P$, deterministic input $x$,
configurations $C$, and optional reference $r$. BMOA (1) builds a strict variant and
candidate variants, (2) executes them and collects numerical and code-generation
evidence, (3) attributes baseline, mechanism, and outcome, and (4) emits an
auditable record.

\subsection{Evidence Collection and Comparisons}

For every configuration, BMOA records the scalar output, raw
\texttt{float32} bits, compiler identity and command, input identifier, and
generated assembly. Repeated executions test stability. When available, a
higher-precision implementation evaluates the same expression over the exact
represented \texttt{float32} inputs. BMOA computes absolute and relative
differences, ULP distance, and exceptional-value mismatches (NaN, infinity, and
signed zero). Shadow and higher-precision execution are established techniques
for exposing numerical error \cite{benz2012dynamic,courbet2021nsan,
chowdhary2021shadow,chowdhary2022eft}; here they supply outcome evidence rather
than a compiler-correctness verdict.

For mechanism $m$, paired configurations $c_m^+$ and $c_m^-$ yield
\begin{equation}
  D_m(x)=d\bigl(y_{c_m^+}(x),y_{c_m^-}(x)\bigr).
\end{equation}
The evidence bundle additionally records
whether assembly contains mechanism-consistent patterns such as FMA or vector
instructions. Assembly corroborates an attribution but cannot establish that a
particular instruction caused the final deviation.

\subsection{Attribution Rules and Records}

For each evaluated comparison, BMOA first emits a separate record for every
baseline being queried. This matters because the same candidate execution can
support a strict-FP record, a numerical-reference record, and a
mechanism-local record without those records making the same claim. BMOA then
assigns:
\begin{itemize}[label=\tiny$\blacksquare$, leftmargin=0pt]

  \item Baseline ($B$). \emph{strict FP}, \emph{numerical reference},
  \emph{mechanism local}, \emph{reproducibility}, or \emph{across compiler},
  according to the comparison being interpreted.
  \item Mechanism ($M$). The unique supported category when one controlled pair
  differs; \emph{mixed}, \emph{ambiguous}, or \emph{unknown} otherwise. A
  compound flag is never promoted automatically to a single-pass cause.
  \item Outcome ($O$). \emph{improvement} if $E_c<E_s$, \emph{degradation} if
  $E_c>E_s$, \emph{neutral} if $E_c=E_s$, and \emph{unknown} when no suitable
  finite reference is available.

\end{itemize}

Formally, for candidate $c$ and input $x$, the supported-mechanism set is
\begin{equation}
S(c,x)=\{m\mid D_m(x)>\tau_m\ \land\
\mathcal{E}_m(c,x)\text{ is consistent with }m\}.
\label{eq:support}
\end{equation}
A singleton $S$ yields its sole category; multiple independently controlled
categories yield \emph{mixed}. A changed compound configuration whose
constituents cannot be separated yields \emph{ambiguous}, whereas an empty $S$
yields \emph{unknown}. The distinction prevents ``flag enabled'' from being
reported as ``mechanism caused the difference.''

The current evaluation sets $\tau_m=0$ and uses exact bit inequality for finite
\texttt{float32} results. It orders finite reference errors with zero tolerance.
NaN, infinities, and signed zeros are reported explicitly; a non-finite case
does not receive an improvement or degradation label unless the selected
reference defines a meaningful ordering. Each record
$R=\langle B,M,O,\mathcal{N},\mathcal{C},\mathcal{E}\rangle$ retains numerical
metrics $\mathcal{N}$, test context $\mathcal{C}$, and supporting evidence
$\mathcal{E}$. One execution can participate in several baseline-specific
records, so record counts measure diagnostic coverage rather than independent
samples.

In Figure~\ref{fig:bmoa-overview}, the strict output is $61$, the candidate is
$992$, and the Decimal reference is $1952$. The candidate has a strict-FP
deviation of $931$ and the reassociation pair changes the output, but its
reference error is $960$ rather than $1891$. The resulting claim is therefore
\emph{strict-FP deviation, reassociation supported, accuracy improvement}; none
of its three components follows from the other two.

\subsection{Evidence-Combination Procedure}

BMOA applies the three attribution decisions independently and in a fixed
order. First, it verifies that the candidate and all comparators consumed the
same serialized input and records their compiler metadata. Second, it creates
one baseline-specific comparison for each diagnostic question being asked.
Third, it evaluates every available controlled mechanism pair and constructs
$S(c,x)$ from Eq.~\ref{eq:support}; no mechanism is selected merely because a
flag or instruction is present. Fourth, it computes $E_s$ and $E_c$ against the
named numerical reference and assigns the outcome from their ordering. Finally,
it attaches the commands, output bits, metrics, and evidence identifiers needed
to replay each decision. 

\begin{table}[t]
  \caption{Evidence required for each diagnostic claim.}
  \label{tab:claim-evidence}
  \centering
  \small
  \begin{tabular}{@{}L{0.69in}L{1.18in}L{1.04in}@{}}
    \toprule
    Claim & Required comparison & Not implied by the claim \\
    \midrule
    Baseline deviation & Candidate versus named baseline & Accuracy or cause \\
    Mechanism support & Isolated pair plus consistent evidence & Semantic violation \\
    Accuracy outcome & Two errors to the same reference & Mechanism or severity \\
    Stability & Repeated identical binary and input & Cross-platform agreement \\
    Compiler specificity & Matched cross-compiler configurations & Transformation cause \\
    \bottomrule
  \end{tabular}
\end{table}

The procedure enforces four record invariants. \emph{Baseline independence}
stores strict-FP, reference, reproducibility, and cross-compiler comparisons in
separate records. \emph{Outcome independence} determines accuracy through the
numerical reference, independently of the sign or magnitude of the strict-FP
deviation. \emph{Conservative mechanism support} retains mixed, ambiguous, and
unknown labels according to the available evidence. \emph{Evidence
traceability} links every label to the paired outputs and configuration
metadata that support it. The strength of each field therefore reflects the
evidence available for that specific comparison.

This ordering also handles control cases consistently. When an ablation changes
the generated code while preserving the output bits, the mechanism-local record
reports zero observed numerical effect. A candidate that produces identical
results across repeated executions receives a stable reproducibility record,
independently of its strict-FP deviation. Equal reference errors produce a
neutral outcome, independently of any bit-level difference between the
candidate and strict-FP outputs. These control records distinguish baseline or
mechanism coverage from evidence of an observed numerical effect.

\subsection{Implementation and Artifact Traceability}

The prototype orchestrates source generation, compilation, execution, reference
evaluation, and record construction. Inputs are serialized once and reused so
that every paired configuration receives identical \texttt{float32} bit
patterns. Evidence is stored across linked artifact tables rather than embedded
in a single record. Attribution records contain case identifiers, output bits,
reference values, and comparison metrics; build logs contain compiler paths,
versions, flags, and build status; environment metadata records the platform,
seed, configurations, and case definitions; and assembly-evidence tables record
generated-assembly paths and instruction-pattern counts. Compiler,
configuration, kernel, and case identifiers, where applicable, connect these
artifacts to raw-output, repeated-run, numerical-reference, and cross-compiler
comparisons. The linked artifacts make the origin of each reported result
auditable and provide the evidence needed to reapply the attribution rules.

For the controlled mechanism matrix, the tables retain evidence at the
granularity at which it is produced. Raw-output and attribution rows are keyed
by compiler, configuration, kernel, and case; they store the scalar value, bit
pattern, selected references, paired comparator, and numerical differences.
Reproducibility rows retain the variation across the three executions, while
cross-compiler rows match corresponding configurations and cases across the two
Clang distributions. Build and assembly rows are keyed by compiler,
configuration, and kernel because one generated binary and its assembly serve
all relevant inputs. This separation avoids copying configuration-level
provenance into every numerical row while retaining explicit join keys.

The controlled run materializes 162 raw-output, attribution, baseline-
consistency, and reproducibility rows, one per configuration instance. It also
stores 81 matched cross-compiler rows, 18 higher-precision baseline rows, and
108 build and assembly rows. The latter count reflects two distributions,
nine configurations, and six kernels; the case-specific tables then associate
each usable binary with its designated inputs. These row types are reported
separately because they represent different comparisons and experimental
granularities, not independent samples.

A controlled-matrix result can be audited by starting from its case identifier
and serialized input, retrieving the strict and candidate raw outputs, and
recomputing their bit, absolute, and ULP differences. The
\texttt{source\_pair\_baseline} field identifies the paired configuration used
for mechanism-local evidence. The stored Decimal, NumPy, and stable-reference
values permit the corresponding error orderings to be recomputed; separate
tables supply repeated-run and cross-compiler checks. Finally, the build log and
assembly-evidence table expose the compiler flags and corroborating instruction
patterns. This replay audits the category-level attribution from its recorded
comparisons; it does not establish instruction-level causality.

Mechanism evidence is asymmetric by design. A nonzero paired comparison is
required to support an output-affecting mechanism. An instruction scan is
only corroborating evidence. Conversely, unchanged output bits do not imply
unchanged code: they record a negative numerical result for that test. The
prototype preserves both positive and negative records because the latter
distinguish coverage from observed effect and expose cases in which code
generation changes without changing the measured output.

\section{Experimental Design}
\label{sec:design}

The evaluation treats BMOA as a diagnostic workflow, not a performance
benchmark or compiler-bug oracle. It has two coordinated components. The
\emph{attribution corpus} measures baseline-specific diagnostic coverage over
five common kernels. The \emph{controlled mechanism matrix} applies mechanism
ablations and robustness checks to nine kernel--input pairs drawn from six
kernels. It addresses four questions:
\begin{itemize}[label=\tiny$\blacksquare$, leftmargin=0pt]

  \item RQ1 (baseline). Does the selected baseline change the diagnosis of the
  same execution results?
  \item RQ2 (mechanism/outcome). Which mechanism categories are supported, and
  how do their deviations affect reference-relative accuracy?
  \item RQ3 (input). Which stress conditions expose or amplify the supported
  mechanism categories?
  \item RQ4 (robustness). Are labels stable under repeat execution, an
  alternative numerical reference, and a second Clang distribution?

\end{itemize}

\paragraph{Kernels and inputs.}
We use six small C kernels. Reduction, dot product, polynomial evaluation,
reciprocal summation, and threshold branching form the attribution corpus; a
purpose-built FMA accumulation kernel isolates contraction in the controlled
mechanism matrix. Deterministic families cover uniform, large-dynamic-range,
near-canceling, mixed-magnitude, near-threshold, and near-zero-denominator
values. Such targeted generation is standard for finding rare floating-point
failures \cite{yu2023inputs,miao2024ranges,yi2024fpcc}; the families expose
sensitivity but do not model workload prevalence. All inputs are generated
deterministically and reused unchanged across compiler configurations.

\begin{table}[t]
  \caption{Kernels and targeted numerical sensitivities.}
  \label{tab:kernels}
  \centering
  \small
  \begin{tabular}{@{}ll@{}}
    \toprule
    Kernel & Primary stress conditions \\
    \midrule
    Reduction & range, cancellation, mixed magnitude \\
    Dot product & range, cancellation \\
    Polynomial & cancellation \\
    Reciprocal sum & near-zero denominator \\
    Threshold branch & near-threshold values \\
    FMA accumulation & multiply-add cancellation \\
    \bottomrule
  \end{tabular}
\end{table}

The small kernels isolate rounding-order, contraction, reciprocal, and
control-flow effects without application-level confounders. The controlled
mechanism matrix contains nine kernel--input pairs: three reduction cases, two
dot-product cases, and one case for each of polynomial evaluation, reciprocal
summation, threshold branching, and FMA accumulation. This mechanism-oriented
design improves attribution interpretability but is not intended to represent
the distribution of real scientific workloads.

\paragraph{Configurations.}
The attribution corpus uses Apple clang~17 on ARM64. The controlled mechanism
matrix applies the nine exact configurations in Table~\ref{tab:configs} using
both Apple clang~17 and Homebrew clang~22 on the same machine. Both are
Clang/LLVM distributions.
Reassociation and
fast math are compound interventions and can support ambiguous or mixed labels.

\begin{table*}[t]
  \caption{Controlled configurations. All commands use \texttt{-O3}; entries
  show the remaining flags. ``Pair'' is the mechanism-local comparator.}
  \label{tab:configs}
  \centering
  \scriptsize
  \begin{tabular}{@{}lll@{}}
    \toprule
    Configuration & Remaining flags & Pair / role \\
    \midrule
    Strict FP & \texttt{-fno-fast-math -ffp-contract=off} & operational baseline \\
    No vectorization & \texttt{-fno-fast-math -ffp-contract=off -fno-vectorize -fno-slp-vectorize} & strict FP \\
    Vectorization & \texttt{-fno-fast-math -ffp-contract=off -fvectorize -fslp-vectorize} & no vectorization \\
    FMA contraction & \texttt{-fno-fast-math -ffp-contract=fast} & strict FP \\
    Reassociation & \texttt{-fassociative-math -fno-signed-zeros -fno-trapping-math -ffp-contract=off} & strict FP; compound \\
    Reciprocal math & \texttt{-freciprocal-math -ffp-contract=off} & strict FP \\
    No signed zero & \texttt{-fno-signed-zeros -ffp-contract=off} & strict FP \\
    Finite math & \texttt{-ffinite-math-only -ffp-contract=off} & strict FP \\
    Fast math & \texttt{-ffast-math -ffp-contract=fast} & strict FP; compound \\
    \bottomrule
  \end{tabular}
\end{table*}

\paragraph{References, metrics, and scale.}
The attribution corpus uses NumPy \texttt{float64} as a scalable
higher-precision proxy for \texttt{float32} outputs, not an exact oracle. The
controlled mechanism matrix uses Decimal with 90 digits over exact
\texttt{float32} inputs and, where applicable, a stable \texttt{float32}
algorithmic reference. Metrics are raw-bit mismatch, absolute and relative
error, ULP distance, and exceptional-value mismatch.

Reference construction preserves the intended input semantics: each Decimal
operand is created from its exact represented \texttt{float32} value. NumPy
widens the same inputs to \texttt{float64}. For reductions, the stable
alternative changes the finite-precision algorithm and therefore addresses a
different question from Decimal. Differences between their results are recorded
as reference sensitivity. ULP distance measures representational separation,
while absolute and relative errors provide the primary severity measures,
especially across sign changes and near zero.

The attribution corpus contains 1,276 baseline-specific BMOA records. The
controlled mechanism matrix contains $2$ distributions $\times$ $9$
kernel--input pairs $\times$ $9$ configurations $=162$ configuration
instances. It yields 126 raw mechanism-ablation rows (63 after collapsing
identical distribution outcomes), 81 paired distribution comparisons, 108
compiled binaries, and 486 executions because each configuration instance runs
three times. We report raw records and collapsed combinations separately.

The experimental unit for an outcome claim is a kernel-input-configuration
combination. The two compiler distributions produce duplicate evidence when
their output bits and labels agree, so frequency claims use the 63 collapsed
mechanism combinations. Raw record counts remain useful for describing
pipeline coverage, but they are not treated as statistically independent
observations. All experimental data and scripts required to reproduce the
results will be publicly released upon acceptance.

\section{Results}
\label{sec:results}

\begin{figure*}[t]
  \centering
  \includegraphics[width=\textwidth]{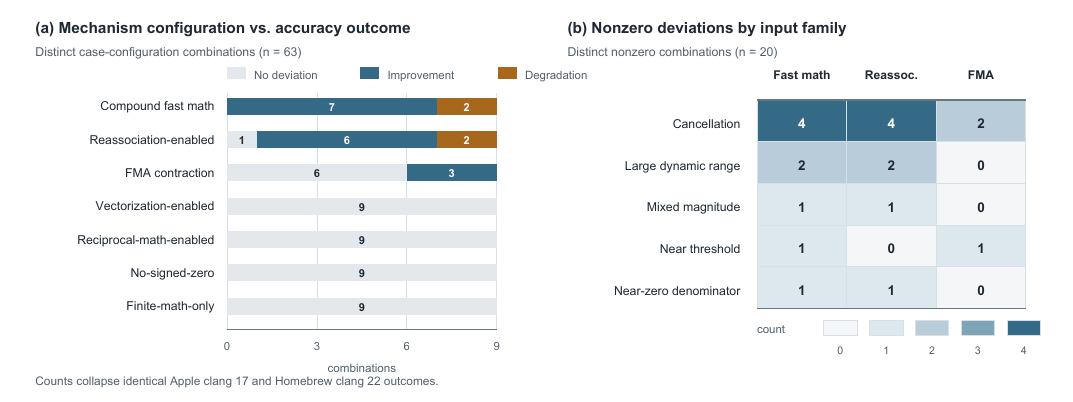}
  \caption{Controlled mechanism-matrix results after collapsing identical
  Apple clang~17 and Homebrew clang~22 outcomes. (a) Outcome counts by
  mechanism configuration; (b) nonzero deviations by input family. Counts are distinct
  kernel-input-configuration combinations.}
  \Description{Stacked bars show 63 mechanism combinations and a heatmap shows the 20 nonzero deviations across input families.}
  \label{fig:results}
\end{figure*}

\subsection{RQ1: Baselines Change the Diagnosis}

The attribution corpus comprises 1,276 records: 375 numerical-reference
comparisons, 225 strict-FP comparisons, 225 mechanism-local comparisons, 375
reproducibility comparisons, and 76 unknown or unsupported attributions. These
are comparison records, not independent failures. The running example demonstrates the
central separation: candidate and strict outputs differ, the ablation supports
reassociation, and the candidate is nevertheless closer to Decimal. Conversely,
a candidate can match strict FP while both have substantial reference error.
Therefore a pass/fail result cannot substitute for the named baseline.

\begin{table}[t]
  \caption{Attribution-corpus BMOA records by diagnostic role.}
  \label{tab:main-records}
  \centering
  \small
  \begin{tabular}{@{}lrl@{}}
    \toprule
    Record class & Count & Question answered \\
    \midrule
    Numerical reference & 375 & Is the result accurate? \\
    Strict FP & 225 & Did the evaluation path change? \\
    Mechanism local & 225 & Which category is supported? \\
    Reproducibility & 375 & Is the binary stable? \\
    Unknown/unsupported & 76 & Where is evidence insufficient? \\
    \midrule
    Total & 1,276 & \\
    \bottomrule
  \end{tabular}
\end{table}

Table~\ref{tab:main-records} also clarifies why records are not collapsed into
one label per execution. A single candidate can simultaneously have a positive
strict-FP record, a negative reproducibility record, an improvement outcome,
and a supported mechanism-local record. The same schema also represents the
opposite combinations: degradation instead of improvement, reference error
without a strict-FP deviation, or a deviation for which the available
ablations support no mechanism. Keeping these observations separate makes an
unknown attribution local to $M$; it does not discard established baseline or
outcome evidence.

\subsection{RQ2--RQ3: Mechanisms, Outcomes, and Inputs}

The 126 raw mechanism-matrix ablation rows contain 40 nonzero deviations: 32
improvements and 8 degradations relative to Decimal. Every result is duplicated
across the two Clang distributions. After collapsing them, 20 of 63 distinct
case-configuration combinations deviate: 16 improve and 4 degrade. As
Figure~\ref{fig:results}(a) shows, compound fast math produces 7 improvements
and 2 degradations; reassociation produces 6 and 2, with one no-deviation case;
and FMA contraction produces 3 improvements and 6 no-deviation cases.
Vectorization, reciprocal math, no-signed-zero, and finite-math configurations
produce no output deviation for these inputs. These negative results do not
generalize beyond the tested matrix.

\begin{table}[t]
  \caption{Raw and cross-distribution-collapsed ablation results.}
  \label{tab:collapse}
  \centering
  \small
  \begin{tabular}{@{}lrrrr@{}}
    \toprule
    Analysis level & Rows & Nonzero & Improve & Degrade \\
    \midrule
    Raw (two distributions) & 126 & 40 & 32 & 8 \\
    Distinct combinations & 63 & 20 & 16 & 4 \\
    \bottomrule
  \end{tabular}
\end{table}

The collapse in Table~\ref{tab:collapse} is an analysis correction, not an
increase in confidence from replication: all 63 Apple/Homebrew pairs agree, so
counting both would double every observed outcome. The 20/63 proportion is also
not an estimate of compiler-wide prevalence. The matrix intentionally
over-samples sensitive inputs and uses one representative instance for each
kernel--input family pairing; its purpose is to test whether BMOA distinguishes
diagnoses under controlled effects.

Input context separates where effects appear. Cancellation accounts for 10 of
the 20 distinct nonzero combinations and exposes all three active categories.
Large dynamic range contributes four, mixed magnitude two, near threshold two,
and near-zero denominator two. The largest absolute deviation,
$5.76\times10^{17}$, occurs for reassociation on a large-range reduction and is
an accuracy degradation; a large-range dot-product deviation of
$3.52\times10^{13}$ is an improvement. The FMA cancellation case reaches a ULP
distance of $2.19\times10^9$ while improving reference accuracy. Mechanism and
magnitude therefore do not determine outcome.

The per-mechanism totals sharpen this result. Compound fast math accounts for
9 of the 20 distinct deviations, reassociation for 8, and FMA contraction for
3. All three FMA deviations improve Decimal-relative accuracy in this matrix,
whereas both compound fast math and reassociation contain improvements and
degradations. Thus even within one configuration family, the mechanism label
does not fix $O$. The result is descriptive of deliberately sensitive cases,
not a ranking of optimization safety: the experiment does not sample mechanisms
or scientific workloads according to their frequency in deployed software.

The 43 no-deviation combinations are controls: 36 from the four inactive
families, six FMA, and one reassociation. They show coverage even when paired
edges are zero; the 16--4 outcome split covers only observed deviations.

Input distribution further separates trigger from consequence. Cancellation
produces 4 compound-fast-math, 4 reassociation, and 2 FMA deviations, the only
family to activate every output-changing category. Large range produces two
compound and two reassociation deviations; each remaining family produces two.
These counts locate evidence for $M$, while outcome still requires the reference
comparison and importance requires an application tolerance.

\subsection{Representative Attributions}

Table~\ref{tab:cases} reports five distinct deviations, with identical Apple
and Homebrew observations combined into a single entry. These cases show the
value of retaining multiple numerical metrics in each BMOA record. The two
large-range cases have the same ULP distance, while their absolute deviations
differ by four orders of magnitude and their outcomes have opposite
directions. The FMA cancellation case has the largest ULP distance in the table
and remains closer to the Decimal reference.

\begin{table*}[t]
  \caption{Representative distinct deviations. The baseline is strict FP in
  every row; outcome is relative to Decimal.}
  \label{tab:cases}
  \centering
  \small
  \begin{tabular}{@{}llllrr@{}}
    \toprule
    Kernel & Input & Supported mechanism & Outcome & Absolute deviation & ULP distance \\
    \midrule
    Reduction & Large dynamic range & Reassociation & Degradation & $5.76\times10^{17}$ & $8$ \\
    Dot product & Large dynamic range & Reassociation & Improvement & $3.52\times10^{13}$ & $8$ \\
    Reduction & Mixed magnitude & Reassociation & Improvement & $931$ & $3.38\times10^7$ \\
    FMA accumulation & Cancellation & FMA contraction & Improvement & $53.69$ & $2.19\times10^9$ \\
    Reduction & Cancellation & Reassociation & Degradation & $207.5$ & $106{,}240$ \\
    \bottomrule
  \end{tabular}
\end{table*}

\paragraph{Mixed magnitude.}
The running example changes from $61$ to $992$ and has a large ULP distance
because the result is small compared with the magnitudes being reduced. The
paired reassociation comparison supports $M=\text{reassociation}$, while the
Decimal errors $1891$ and $960$ establish $O=\text{improvement}$. Reporting
only the strict-FP mismatch would incorrectly suggest that the optimization
necessarily made the result worse.

\paragraph{Large range and cancellation.}
Reassociation in the large-range reduction moves the output by
$5.76\times10^{17}$ but only eight ULPs and degrades the Decimal-relative
result. The corresponding dot product also moves eight ULPs yet improves
accuracy. In the FMA case, contraction changes where rounding occurs; the
paired configuration and instruction evidence support the category, but the
outcome still comes exclusively from reference-error ordering. These cases
show that baseline, mechanism, absolute deviation, ULP distance, and outcome
are complementary fields, not interchangeable severity labels.

\paragraph{Negative mechanism tests.}
The vectorization-enabled binaries contain vector instructions in many assembly
files, but their outputs match the no-vectorization comparator for all nine
kernel--input pairs in the controlled matrix. BMOA records the code-generation evidence and the
zero-deviation outcome without claiming a numerical vectorization effect. This
negative case is important: instruction presence is insufficient for numerical
causality, while absence of an observed effect is bounded to the tested inputs.

\subsection{RQ4: Validation Bounds the Claims}

\begin{table}[t]
  \caption{Controlled mechanism-matrix validation checks.}
  \label{tab:validation}
  \centering
  \small
  \begin{tabular}{@{}lr@{}}
    \toprule
    Check & Result \\
    \midrule
    Decimal--NumPy outcome agreement & 162/162 \\
    Decimal--stable-reference disagreement & 20/162 \\
    Repeated-run variation & 0/162 \\
    Apple--Homebrew Clang deviation & 0/81 \\
    Assembly files scanned & 108 \\
    FMA / vector evidence rows & 16 / 102 \\
    Reciprocal-estimate evidence rows & 0 \\
    \bottomrule
  \end{tabular}
\end{table}

Decimal and NumPy assign the same outcome to all 162 mechanism-matrix instances,
supporting NumPy for this test matrix but not establishing a general oracle.
Decimal and the stable \texttt{float32} algorithm disagree on 20 labels because
they answer different numerical questions; BMOA records this reference
sensitivity. All repeated executions and 81 paired Clang-distribution
comparisons are bit-identical. Thus nondeterminism and distribution choice do
not explain the observed deviations, but the study contains no positive case
for either diagnostic category.

The 162 configuration instances comprise two distributions, nine kernel--input
cases, and nine configurations. Three runs per instance yield 486 executions;
bit equality within each triplet supplies 162 negative reproducibility results.
Matching the distributions yields 81 cross-compiler pairs. Their equality
justifies collapsing duplicates, but not agreement for other compilers or
targets: both are Clang/LLVM distributions on the same machine.

The 20 Decimal--stable disagreements reduce to ten distinct combinations after
collapsing identical compiler-distribution outcomes: four compound-fast-math,
three FMA-contraction, and three reassociation combinations. They occur in
cancellation, mixed-magnitude, and near-threshold cases; no disagreement occurs
for the four inactive configuration families. In every disagreement, Decimal
ranks the candidate as an improvement whereas the stable \texttt{float32}
algorithm ranks it as a degradation. This consistent reversal reflects their
different specifications rather than reference noise: Decimal evaluates the
real-arithmetic expression over exact represented inputs, while the stable
reference changes the finite-precision algorithm. The record must therefore
retain both the reference identity and its outcome.

Across 108 assembly files, vector patterns occur in 102 and FMA patterns in 16,
yet output changes occur in no vectorization case and only three collapsed FMA
cases. No reciprocal-estimate pattern is found. These gaps establish
instruction presence as corroborating rather than causal evidence. Overall,
the results support BMO separation within the tested kernels, inputs,
references, configurations, and ARM64 environment.

\section{Discussion and Limitations}
\label{sec:discussion}

BMOA is best understood as a diagnostic report contract and an empirical
interface to later formal reasoning about compiler-aware numerical correctness.
It prevents three conflations: reference error is not automatically
compiler-induced, strict-FP deviation is not automatically an accuracy
regression, and a deterministic configuration difference is not run-to-run
variation. For later formalization, $B$ names the comparison relation and
system boundary, $M$ identifies compiler behavior whose assumptions may require
modeling, and $O$ names the reference-qualified numerical objective. The
evidence bundle supplies the inputs, configurations, outputs, and metrics from
which downstream proof obligations can be formulated. Unknown and ambiguous
labels expose missing models or evidence rather than inventing a cause. BMOA
itself neither proves semantic equivalence nor certifies an accuracy bound.

The attribution remains operational and category-level. Compiler flags may
activate overlapping transformations, phase ordering may couple them, and
instruction presence provides corroborating evidence without establishing an
effect on the final output. Strict FP serves as a controlled comparator, while
formal semantic correctness requires tools such as LifeJacket or verified
fast-math reasoning \cite{notzli2016lifejacket,becker2019icing}. LLVM
optimization records, IR differencing, pass disabling, dynamic provenance, and
numerical localizers could strengthen $\mathcal{E}$ and support downstream
proof obligations while preserving the BMO dimensions.

Outcome is relative to a named reference and tolerance, and zero-tolerance
ordering is not application severity. The empirical scope is also deliberately
narrow: small kernels, targeted stress inputs, one seed, one ARM64 CPU, and two
Clang distributions whose outcomes are identical. Paired ablations, serialized
inputs, exact-bit comparisons, repetitions, and Decimal validation improve
internal confidence but do not establish population frequencies. GCC, x86-64,
GPUs, parallel reductions, math libraries, applications, multiple seeds, MPFR
or interval references, and positive nondeterminism and cross-compiler cases
remain necessary tests of generality.

\section{Related Work}
\label{sec:related}

\paragraph{Compiler variability and testing.}
FLiT, pLiner, Ciel, and Varity detect, quantify, or localize variability from
functions to expressions and across heterogeneous platforms
\cite{sawaya2017flit,guo2020pliner,miao2023ciel,laguna2020varity}; recent work
generates triggering values and ranges \cite{yu2023inputs,miao2024ranges}.
Csmith and equivalence-modulo-inputs expose compiler defects by differential or
metamorphic testing \cite{yang2011csmith,le2014emi}. BMOA is complementary: it
starts with an observed difference and attributes its meaning, supported
mechanism category, and accuracy outcome.

\paragraph{Numerical diagnosis and testing.}
FpDebug, Herbgrind, NSan, and shadow-execution systems detect and localize error
with higher-precision or error-free computations
\cite{benz2012dynamic,sanchezstern2018herbgrind,courbet2021nsan,
chowdhary2021shadow,chowdhary2022eft}. FPChecker detects GPU floating-point
exceptions, while FPDiff and empirical studies expose discrepancies and
real-world numerical bug patterns
\cite{laguna2019fpchecker,vanover2020fpdiff,difranco2017bugs}. BMOA consumes
such evidence but focuses specifically on interpretation across controlled
compiler configurations.

\paragraph{Analysis, transformation, and reproducibility.}
Herbie improves expressions; Daisy and FPTaylor bound roundoff; backward-error
analysis and tool composition provide complementary guarantees
\cite{panchekha2015herbie,darulova2018daisy,solovyev2018fptaylor,
fu2015backward,becker2018combining}. Precimonious tunes precision, Verificarlo
samples rounding sensitivity, and reproducible summation controls reduction
order \cite{rubio2013precimonious,denis2016verificarlo,demmel2013repro,
collange2015repro}. Repair synthesis changes erroneous programs
\cite{zou2022repair}; BMOA instead leaves the program unchanged and produces an
evidence-bounded diagnostic record.

\section{Conclusion}
\label{sec:conclusion}

We propose BMOA, a diagnostic workflow for analyzing compiler-induced
floating-point deviations. BMOA separates each deviation into three distinct
claims: the baseline that defines its meaning, the mechanism category supported
by controlled evidence, and the numerical outcome relative to a reference.
Across a 1,276-record attribution corpus and a 162-instance controlled
mechanism matrix, BMOA identifies both accuracy improvements and degradations.
Within the tested matrix, cancellation
and large dynamic range produce the strongest amplification. BMOA transforms
raw numerical mismatches into explicit and auditable diagnostic claims whose
scope is defined by the tested baselines, configurations, inputs, and
references. By making comparison relations and evidence obligations explicit,
BMOA provides an empirical foundation for future formal specifications and
proof obligations for compiler-aware numerical correctness.

\bibliographystyle{ACM-Reference-Format}
\bibliography{sample-base}

\end{document}